\documentclass[11pt]{article}
\usepackage{verbatim} 
\usepackage[preprint]{acl}
\usepackage{booktabs}
\usepackage{tabularx}
\usepackage{times}
\usepackage{latexsym}
\usepackage{graphicx} 
\usepackage{subfigure}
\usepackage{float}
\usepackage[linesnumbered,ruled,vlined]{algorithm2e}
\usepackage{amsmath}  %
\usepackage{makecell} %
\usepackage[T1]{fontenc}

\usepackage[utf8]{inputenc}

\usepackage{microtype}

\usepackage{inconsolata}

\usepackage{graphicx}

\usepackage{etoc}
\usepackage{scrextend}
\title{
HopRAG: Multi-Hop Reasoning for Logic-Aware \\
Retrieval-Augmented Generation
}

\author{Hao Liu$^{1\dagger}$, Zhengren Wang$^{1,2\dagger}$, Xi Chen$^{3}$, Zhiyu Li$^{2*}$,\\{\bf Feiyu Xiong$^{2}$}, {\bf Qinhan Yu$^{1}$}, {\bf Wentao Zhang$^{1*}$} \\
$^{1}$Peking University $^{2}$Center for LLM, Institute for Advanced Algorithms Research, Shanghai \\
$^{3}$Huazhong University of Science and Technology \\
\texttt{\{liuhao\_2002,wzr,yuqinhan\}@stu.pku.edu.cn} ~~~~~ \texttt{xichenai@hust.edu.cn} \\ \texttt{\{lizy, xiongfy\}@iaar.ac.cn} ~~~~~ \texttt{wentao.zhang@pku.edu.cn} 
}

\begin{document}
\maketitle
\begingroup
\deffootnote[1.5em]{1.5em}{1em}{}
\renewcommand\thefootnote{}\footnote{
$\dagger$ Equal contribution;  * Corresponding author.
}
\addtocounter{footnote}{-1}
\endgroup

\begin{abstract}
Retrieval-Augmented Generation (RAG) systems often struggle with imperfect retrieval, as traditional retrievers focus on lexical or semantic similarity rather than logical relevance. To address this, we propose \textbf{HopRAG}, a novel RAG framework that augments retrieval with logical reasoning through graph-structured knowledge exploration. During indexing, HopRAG constructs a passage graph, with text chunks as vertices and logical connections established via LLM-generated pseudo-queries as edges. During retrieval, it employs a \textit{retrieve-reason-prune} mechanism: starting with lexically or semantically similar passages, the system explores multi-hop neighbors guided by pseudo-queries and LLM reasoning to identify truly relevant ones. Experiments on multiple multi-hop benchmarks demonstrate that HopRAG’s \textit{retrieve-reason-prune} mechanism can expand the retrieval scope based on logical connections and improve final answer quality.

\end{abstract}

\section{Introduction}
\label{sec:introduction}
\label{introduction}

\begin{quote}
    \textit{``Everyone and everything is six or fewer steps away, by way of introduction, from any other person in the world.''} 
\begin{flushright}
\qquad\;\;\, --- Six Degrees of Separation
\end{flushright}
\end{quote}
Retrieval-augmented generation (RAG) has become the standard approach for large language models (LLMs) to tackle knowledge-intensive tasks \citep{guu2020retrieval,lewis2020retrieval, izacard2022few, min-etal-2023-nonparametric, ram2023context,liang2025saferag}. Not only can it effectively address the inherent knowledge limitations and hallucination issues \cite{zhang_sirens_2023}, but it can also enable easy interpretability and provenance tracking \citep{akyurek-etal-2022-towards}. Especially, the efficacy of RAG hinges on its retrieval module for identifying relevant documents from a vast corpus.

Currently, there are two mainstream types of retrievers: sparse retrievers \citep{TF-IDF,BM25} and dense retrievers \citep{BGE,e52024multilingual,sturua2024jina,wang_qaencoder_2024}, which focus on lexical similarity and semantic similarity respectively, and are often combined for better retrieval performance \citep{sawarkar2024blended}.
Despite advancements, the ultimate goal of information retrieval extends beyond lexical and semantic similarity, striving instead for \textit{logical relevance}. Due to the lack of logic-aware mechanism, the imperfect retrieval remains prominent \citep{wang2024astute, shao2024scaling, dai2024unifying, su2024bright, su2024brightrealisticchallengingbenchmark}. For precision, the retrieval system may return lexically and semantically similar but indirectly relevant passages; regarding recall, it may fail to retrieve all the necessary passages for the user query. 

Both cases eventually lead to inaccurate or incomplete LLM responses \citep{chen2024benchmarking,xiang2024certifiably,zou2024poisonedrag}, especially for multi-hop or multi-document QA tasks requiring multiple relevant passages for the final answer.  
In contrast, the reasoning capability of generative models is rapidly advancing, with notable examples such as OpenAI-o1 \citep{jaech2024openai} and DeepSeek-R1 \citep{guo2025deepseek}. Therefore, a natural research question arises: \textit{"Is it possible to introduce reasoning capability into the retrieval module for more advanced RAG systems?"} 

\begin{figure*}[ht]
\centering
\subfigure[Precision, recall and F1 score]{\label{chunk precision and recall}\includegraphics[width=1\columnwidth]{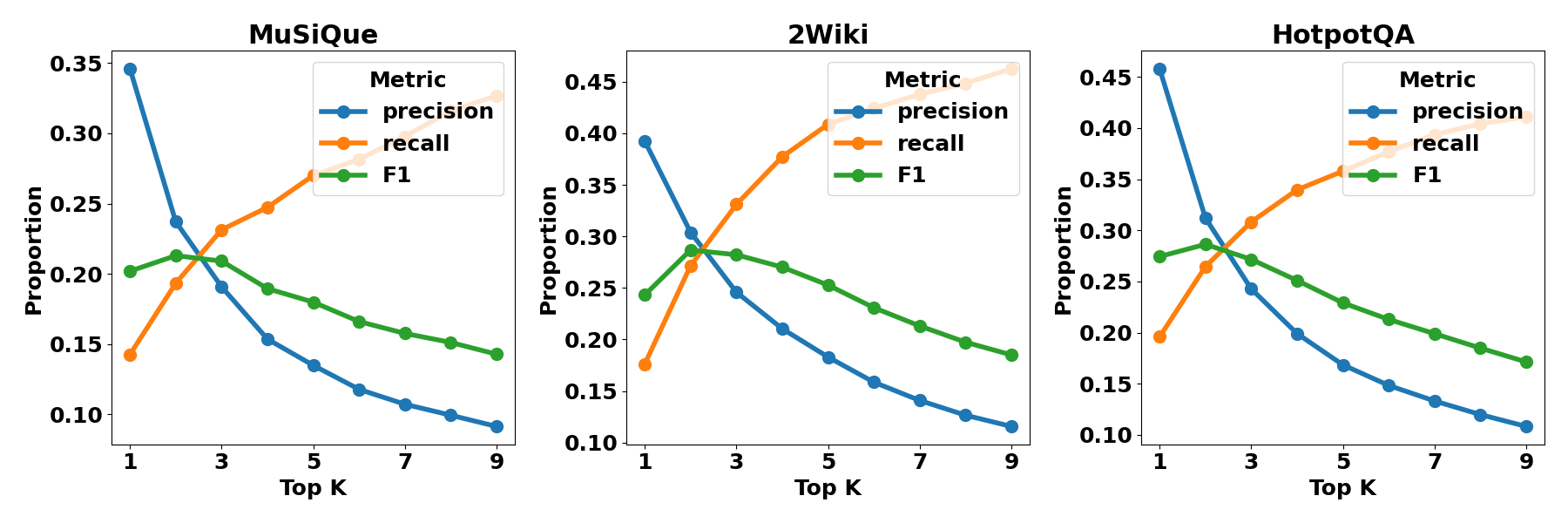}}
\subfigure[Proportions of passages on relevance]
{\label{relevance}\includegraphics[width=1\columnwidth]{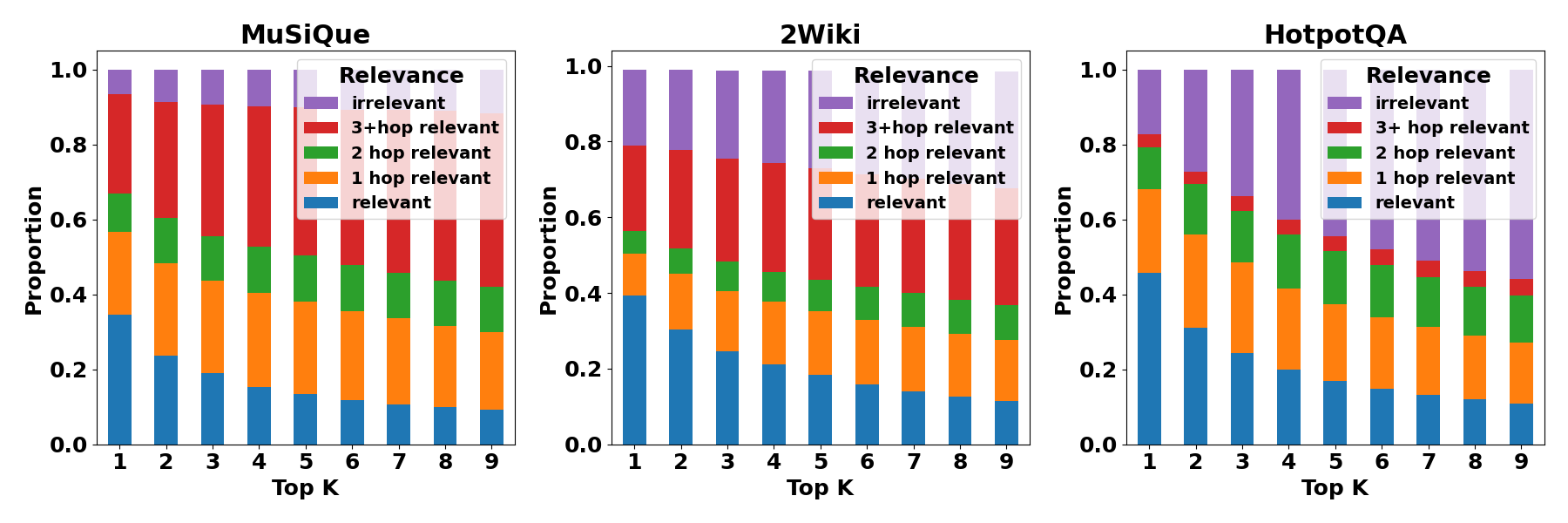}}
\caption{(a) Precision, recall and F1 score of BGE dense retrievers on MuSiQue, 2WikiMultiHopQA and HotpotQA with different $top_k$ parameters, revealing the severe imperfect retrieval phenomenon. The highest recall reaches saturation at 0.45 in our settings.
(b) We categorize retrieved passages into \textbf{relevant}, \textbf{indirectly relevant} and \textbf{irrelevant} according to the logical relevance to the query. The relevant passages are exactly the supporting facts, and indirectly relevant passages can hop to the supporting facts via HopRAG while irrelevant passages cannot. A large proportion of retrieved passages are indirectly relevant. 
}\label{motivation_figure}
\end{figure*}

From a logical structure perspective, existing RAG systems can be mainly categorized into three types: 
\textbf{Non-structured RAG} simply adopts sparse or dense retrievers. The retrieval is only based on keyword matching or semantic vector similarity, but fails to capture the logical relations between user queries and passages.
\textbf{Tree-structured RAG} \citep{sarthi_raptor_2024, chen_walking_2023, fatehkia2024traglessonsllmtrenches} focuses on the hierarchical logic of passages within a single document, but ignores relations beyond the hierarchical structure or across documents. Further, it introduces redundant information across different levels.
\textbf{Graph-structured RAG} \citep{soman2024biomedicalknowledgegraphoptimizedprompt,kang2023knowledgegraphaugmentedlanguagemodels,edge_local_2024,guo_lightrag_2024} models logical relations in the most ideal form by constructing knowledge graphs (KGs) to represent documents, where entities are vertices and their relations are edges. However, the reliance on predefined schemas limits the flexible expressive capability \citep{li_graph_2024}; constructing and updating knowledge graphs are challenging and prone to errors or omissions \citep {edge_local_2024}; the triplet format of knowledge necessitates extra textualization or fine-tuning to improve LLMs' understanding \citep{he_g-retriever_2024}.

\begin{figure}[t]
  \includegraphics[width=\columnwidth]{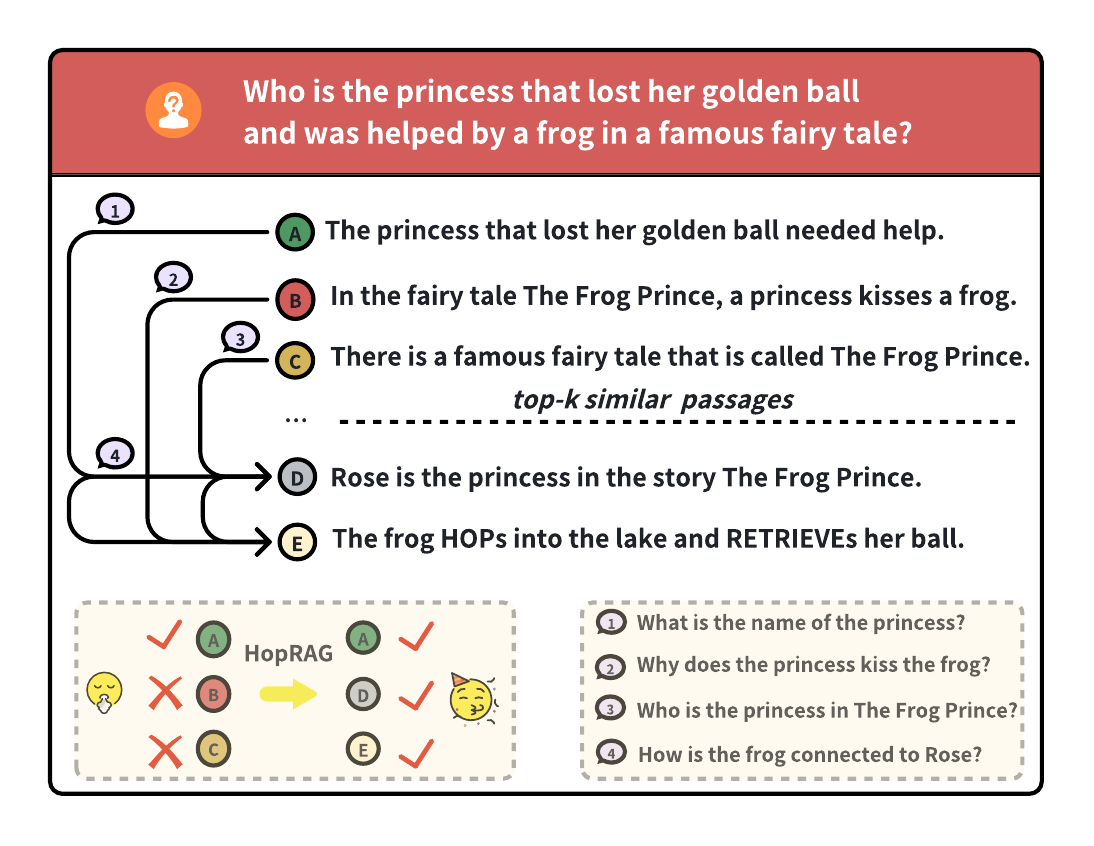}
  \caption{Demonstration of hopping between passages. For the user query, BGE dense retriever can only return one of the three supporting facts within $top_k$ budget. However, lexically or semantically similar passages complement each other.
  Hopping between passages, by questions as pathways, improves the retrieval accuracy and completeness.}
  \label{fig:demo}
\end{figure}

\paragraph{Motivation} As reported by \citep{wang2024astute}, even with advanced real-world search engines, roughly 70\% retrieved passages do not directly contain true answers in their settings. We confirm the severity of imperfect retrieval in terms of both precision and recall, as illustrated in Figure \ref{motivation_figure}(a). Inspired by the small-world theory \citep{kleinberg2000small} or six degrees of separation \citep{guare2016six}, we propose that, \textit{although lexically and semantically similar passages could be indirectly relevant or even distracting, they can serve as helpful starting points to reach truly relevant ones}. As shown in Figure \ref{motivation_figure}(b), considering a graph composed of passages with logical relations as edges, a large proportion of retrieved passages fall within several hops of the ground truths. 

Based on these observations, we propose \textbf{HopRAG}, an innovative graph-structured RAG system. At indexing phase, we construct a graph-structured knowledge index with passages as vertices and logic relations as directed edges. Specifically, the passages are connected by pseudo-queries generated by \textit{query simulation} and \textit{edge merging} operations. For example, as demonstrated in Figure \ref{fig:demo}, the pseudo-query "Why does the princess kiss the frog?" connects the raiser passage and the solver passage, as the pivot for logical hops. During retrieval, we employ reasoning-augmented graph traversal, following a three-step paradigm of retrieval, reasoning, and pruning. This process searches for truly relevant passages within the multi-hop neighborhood of indirectly relevant passages, guided by both the index structure and LLM reasoning.

\paragraph{Contributions}
Our contributions are as follows:
\begin{itemize}
\item We reveal the severe imperfect retrieval phenomenon for multi-hop QA tasks. The results quantify that currently over 60\% of retrieved passages are indirectly relevant or irrelevant. To turn "trash" into "treasure", we further employ indirectly relevant passages as stepping stones to reach truly relevant ones.
\item We propose HopRAG, a novel RAG system with logic-aware retrieval mechanism. As lexically or semantically similar passages complement each other, HopRAG connects the raiser and solver passages with pseudo-queries. Beyond similarity-based retrieval, it reasons and prunes along the queries during retrieval. It also features flexible logical modeling, cross-document organization, efficient construction and updating.
\item Extensive experiments confirm the effectiveness of HopRAG. The retrieve-reason-prune mechanism achieves over 36.25\% higher answer metric and 20.97\% higher retrieval F1 score compared to conventional information retrieval approaches. Several ablation studies provide more valuable insights.
\end{itemize}

\section{Related Work} 
\label{sec:RelatedWork}
\paragraph{Retrieval-Augmented Generation}
Retrieval-augmented generation significantly improves large language models by incorporating a retrieval module that fetches relevant information from external knowledge sources \citep{EaE, REALM, FID, survey, yu2025mramg}. Retrieval models have evolved from early sparse retrievers, such as TF-IDF \citep{TF-IDF} and BM25 \citep{BM25}, which rely on word statistics and inverted indices, to dense retrievers \citep{2020RAG} that utilize neural representations for semantic matching. Advanced methods, such as Self-RAG \citep{Self-RAG} and FLARE~\citep{FLARE} which determine the necessity and timing of retrieval, represent significant developments. However, the knowledge index remains logically unstructured, with each round of search considering only lexical or semantic similarity.

\paragraph{Tree\&Graph-structured RAG}
Tree and graph are both effective structures for modeling logical relations. RAPTOR~\citep{RAPTOR} recursively embeds, clusters, and summarizes passages, constructing a tree with differing levels of summarization from the bottom up.
MemWalker~\citep{chen_walking_2023} treats the LLM as an interactive agent walking on the tree of summarization. SiReRAG~\citep{zhang2024sireragindexingsimilarrelated} explicitly considers both similar and related information by constructing both similarity tree and relatedness tree. 
PG-RAG~\citep{PG-RAG} prompts LLMs to organize document knowledge into mindmaps, and unifies them for multiple documents. 
GNN-RAG~\citep{GNN-RAG} reasons over dense KG subgraphs with learned GNNs to retrieve answer candidates.
For query-focused summarization, GraphRAG~\citep{GraphRAG} builds a hierarchical graph index with knowledge graph construction and recursive summarization. 
Despite advancements, tree-structured RAG only focuses on the hierarchical logic within a single document; graph-structured RAG is costly, time-consuming, and returns triplets instead of plain text. 
In contrast, HopRAG offers a more lightweight and downstream task friendly alternative, with flexible logical modeling, cross-document organization, efficient construction and updating.

\section{Method} 
\label{sec:method}
\begin{figure*}[htbp!]
\centering
  \includegraphics[width=\textwidth]{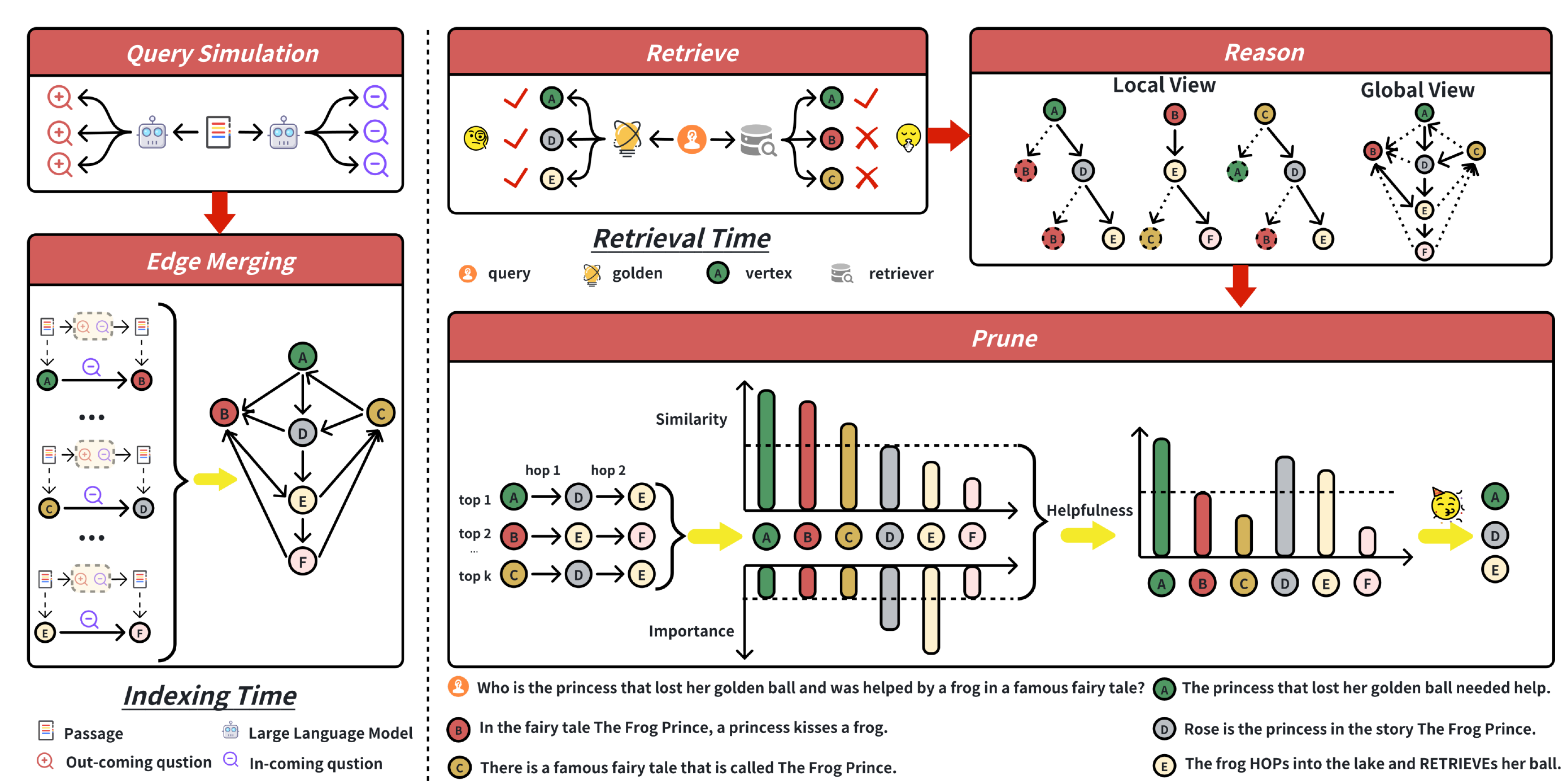}
  \caption{The workflow of HopRAG. \textbf{Left:} At indexing time, we first utilize \textit{Query Simulation} to generate pseudo-queries for each passage and then apply \textit{Edge Merging} to connect passages with directed logical edges. \textbf{Right:} At retrieval time, we employ a \textit{Retrieve-Reason-Prune} pipeline. We first retrieve through purely similarity-based retrieval, then run reasoning-augmented graph traversal to explore the neighborhood, and finally prune the search by a novel metric \textit{Helpfulness} considering both textual similarity and logical importance.}
  \label{fig:overview}
\end{figure*}

In this section, we introduce our logic-aware RAG system, named HopRAG. An overview of this system is illustrated in Figure \ref{fig:overview}.
\subsection{Problem Formulation}

Given a passage corpus $P=\{p_1, p_2, ..., p_N\}$ and a query $q$ which requires the information from multiple passages in $P$, the task is to design (1) a graph-structured RAG knowledge base that not only stores all the passages in corpus $P$ but also models the similarity and logic between passages; (2) a corresponding retrieval strategy that can hop from indirectly relevant passages to truly relevant passages for better retrieval. Finally, with the query $q$ and $k$ passages as context ${C}=\{p_{i_1},p_{i_2},...,p_{i_k}\}$, the LLM generates the response \(\mathcal{O} \sim \mathcal{P}(\mathcal{O}|q, {C}) \).

\subsection{Graph-Structured Index}
We construct a graph-structured index $G=(\mathcal{V},\mathcal{E})$ where the vertex set $\mathcal{V}$ consists of vertices storing all the passages and the directed edge set $\mathcal{E} = \{\langle v_i,e_{i,j},v_j\rangle | v_i,v_j\in \mathcal{V} \}\subset \mathcal{V} \times \mathcal{V}$ is established based on the logical relations between passages for multi-hop reasoning. To establish $G$, we utilize \textbf{Query Simulation} to identify the logical relations and leverage textual similarity for efficient \textbf{Edge Merging}. 
\paragraph{Query Simulation}
To identify the logical relations between passages, we generate a series of pseudo-queries for each passage, and use them to explore the passage's relations with the others and bridge the inherent gap between user-queries and passages \citep{wang_qaencoder_2024}. Specifically, we adopt LLM to generate two groups of pseudo-queries for each passage $p_i$: (1) $m$ out-coming questions $Q_i^+= \bigcup_{1 \le j \le m} \{q_{i,j}^{+} \}$ that originate from this passage but cannot be answered by itself; (2) $n$ in-coming questions $Q_i^-= \bigcup_{1 \le j \le n} \{q_{i,j}^{-}\}$ whose answers are within the passage. As demonstrated in Figure \ref{fig:demo}, for the toy passage "Rose is the princess in the story The Frog Prince", one in-coming question might be "What is the name of the princess?" and one out-coming question might be "How is the frog connected to Rose?". The prompts are in Appendix \ref{prompt}. 

We extract keywords from $Q_i^+$ and $Q_i^-$ using named entity recognition $\text{NER}(\cdot)$ for sparse representation, and embed these questions into semantic vectors using an embedding model $\text{EMB}(\cdot)$ for dense representation. This results in sparse representations $K_i^+= \bigcup_{1 \le j \le m}\{k_{i,j}^{+} \}$ and $K_i^-= \bigcup_{1 \le j \le n}\{k_{i,j}^{-}\}$, and dense representations $V_i^+=\bigcup_{1 \le j \le m} \{\mathbf{v}_{i,j}^{+}\}$ and $V_i^-=\bigcup_{1 \le j \le n} \{\mathbf{v}_{i,j}^{-}\}$. We further define out-coming triplets as $r_{i,j}^+:=(q_{i,j}^{+},k_{i,j}^{+},\mathbf{v}_{i,j}^{+})$ and in-coming triplets $r_{i,j}^-:=(q_{i,j}^{-},k_{i,j}^{-},\mathbf{v}_{i,j}^{-})$. Each passage $p_i$ is stored inside a vertex $v_i$, featured with its out-coming triplets $R_i^+= \bigcup_{1 \le j \le m} \{r_{i,j}^{+}\}$ and in-coming triplets $R_i^-= \bigcup_{1 \le j \le n} \{r_{i,j}^{-} \}$. 

\paragraph{Edge Merging}
Given the out-coming and in-coming triplets, we match paired triplets via hybrid retrieval and establish directed edges between the corresponding passages.
For each out-coming triplet $r_{s,i}^+$ of source vertex $v_s$, the most matching in-coming triplet $r_{t^*,j^*}^-$ is determined as follows:
\begin{equation}
\label{sim}
\begin{aligned}
    \text{SIM}(r_{s,i}^{+}, r_{t,j}^{-}) &= \frac{\frac{|k_{s,i}^{+} \cap k_{t,j}^{-}|}{|k_{s,i}^{+} \cup k_{t,j}^{-}|}+\frac{\mathbf{v}_{s,i}^{+} \cdot \mathbf{v}_{t,j}^{-}}{||\mathbf{v}_{s,i}^{+}|| \cdot ||\mathbf{v}_{t,j}^{-}||}}{2} \\
    r_{t^*,j^*}^- &= \arg\max_{r_{t,j}^{-}  } \text{SIM}(r_{s,i}^{+}, r_{t,j}^{-})
\end{aligned}
\end{equation}

We then build the directed edge $\langle v_s,e_{s,t^*},v_{t^*} \rangle$ with aggregated features, where $e_{s,t^*}:=({q_{t^*,j^*}^{-}},{k_{t^*,j^*}^{-}\cup k_{s,i}^{+}},{\mathbf{v}_{t^*,j^*}^{-}})$. 

\subsection{Reasoning-Augmented Graph Traversal }
For more accurate and complete responses, HopRAG's retrieval strategy leverages the reasoning ability of LLM to explore the neighborhood of probably indirectly relevant passages and hop to relevant ones based on the logical relations in the graph structure. As shown in Algorithm \ref{ragt}, by reasoning over the questions on out edges $e_{i,j}$ of a current vertex $v_{i}$ and then choosing to hop to the most promising vertex $v_{j}$, we realize reasoning-augmented graph traversal for better retrieval performance. 

\paragraph{Retrieval Phase} To start the local search over the graph for query $q$, we first use $\text{NER}(\cdot)$ and $\text{EMB}(\cdot)$ to get the keywords $ k_{q}$ and vector $\mathbf{v}_{q}$ of $q$, which will be used for hybrid retrieval to match $top_k$ similar edges $\langle v_i,e_{i,j},v_j \rangle $, following Equation \ref{sim}. With each vertex $v_j$ from these edges we initialize a context queue $C_{queue}$ for breadth-first local search \citep{voudouris2010guided}. 

\paragraph{Reasoning Phase} To fully exploit the logical relations over the graph and hop from indirectly relevant vertices to relevant ones, we introduce breadth-first local search which utilizes the LLM to choose the most appropriate neighbor for each $v_j$ in $C_{queue}$ to append to the tail of the queue. Specifically, for each $v_j$ in $C_{queue}$ in each round of hop, we leverage LLM to reason over all the questions from its out edges to choose one $e_{j,k}$ with the question which the LLM regards as the most helpful for answering $q$ and append vertex $v_k$ to $C_{queue}$.  
After hopping from each vertex in the current $C_{queue}$ we can expand the context with at most $top_k$ new vertices. From these new vertices we continue the next round of hop. Since different vertices may hop to the same vertex, we believe the vertices with more visits are more important for answering $q$, and use a counter $C_{count}$ to track the number of visits for each vertex and measure its importance. By conducting $n_{hop}$ rounds of hop, we realize reasoning-augmented graph traversal and expand the context length to at most $(n_{hop}+1)\times top_k$.

\paragraph{Pruning Phase} To avoid including too many intermediate vertices during the traversal, we introduce a novel metric Helpfulness $H(\cdot)$  that integrates similarity and logic to re-rank and then prune the traversal counter $C_{count}$. We calculate $H_i$ following Equation \ref{H} for each $v_i$ in $C_{count}$ and keep the $top_k$ vertices with the highest $H_i$, where hybrid textual similarity $\text{SIM}(v_i, q)$ calculates the average lexical and semantic similarity between the passage in $v
_i$ and query $q$ following Equation \ref{sim}; and $\text{IMP}(v_i,C_{count})$ is defined as the normalized number of visits of $v_i$ in $C_{count}$ during traversal following Equation \ref{Imp}. We prune \( C_{count} \) by retaining $top_k$ vertices with the highest \( H \) value, resulting in the final context \( C \).

\begin{equation}
    H_i=\frac{\text{SIM}(v_i, q)+\text{IMP}(v_i,C_{count})}{2}
    \label{H}
\end{equation}
\begin{equation}
    \text{IMP}(v_i, C_{count}) = \frac{C_{count}[v_i]}{\sum_{v_j \in C_{count}} C_{count}[v_j]}
    \label{Imp}
\end{equation}
\begin{algorithm}
  \SetAlgoLined
  \KwIn{$q$, $top_k$, $n_{hop}$, $G$}
  \KwOut{$C$}
 $\mathbf{v}_{q}$ $\leftarrow$ EMB($q$)\;$  k_{q}$ $\leftarrow$ NER($q$)\;
  $C_{queue}\leftarrow$ Retrieve($\mathbf{v}_{q}$, $k_{q}$, $G$)\;
  $C_{count}\leftarrow $Counter$(C_{queue})$\;
  \For{$i \leftarrow 1,2,...,n_{hop}$ }{
  \For{j $\leftarrow 1,2,...,\mid C_{queue}\mid $} {
  $v_j\leftarrow $ $C_{queue}$.dequeue()\;
  $v_k \leftarrow$ Reason($\{\langle v_j,e_{j,k},v_k\rangle\}$)\;
\eIf{$v_k$  not in $C_{count}$}{
$C_{queue}$.enqueue($v_k$)\;
$C_{count}[v_k]\leftarrow1$ \;
      }{
      $C_{count}[v_k]++$ \;
      }
  }
  }
  $C\leftarrow$Prune($C_{count}$, $\mathbf{v}_{q}$, $k_q$, $top_k$)\;
  \Return $C$
  \caption{Reasoning-Augmented Graph Traversal}
  \label{ragt}
\end{algorithm}

\section{Experiments}

\begin{table*}[htb]
    \centering
\resizebox{0.7\textwidth}{!}{
      \begin{tabular}{ccccccccc}
        \toprule
        & \multicolumn{2}{c}{MuSiQue}& \multicolumn{2}{c}{2Wiki}& \multicolumn{2}{c}{HotpotQA}& \multicolumn{2}{c}{Average}\\
        
        Method& EM & F1  & EM & F1  & EM & F1   & EM &F1   \\
        \midrule
        BM25
& 5.80 & 11.00 
 & 27.00 & 31.55  & 33.40 & 44.30 
  & 22.07 &28.95  \\
        BGE
& 11.80 & 18.60  & 27.90 & 30.80  & 38.40 & 50.56   & 26.03 &33.32  \\
 \thead{Query \\ Decomposition}& 21.50 & 31.40 & 43.90& 47.06& 43.60 & 58.94 & 31.10 &40.01 \\
 Reranking& 24.50 & 34.53 & 46.70& 50.89& 47.70 & 62.95 & 34.67 &43.60 \\
 HippoRAG
& 32.60 & 43.78  & \textbf{66.40}& \textbf{74.01}& 59.90 &74.29 
  & 52.97 &64.03 
 \\
 RAPTOR
& 35.30 & 47.47  & 54.90 & 61.20  & 58.10 &72.48 
  & 49.43 &60.38 
 \\
 SiReRAG& \underline{38.90}& \underline{52.08} & 60.40 & 68.20  & \textbf{62.50}&\underline{77.36}  & \underline{53.93}&\underline{65.88} \\ 
        \midrule
        HopRAG&  \textbf{39.10}&  \textbf{53.00} &  \underline{61.60}&  \underline{68.93}&  \underline{61.30}&  \textbf{78.34}  & \textbf{54.00}&\textbf{66.76} \\
        \bottomrule
    \end{tabular}
}
    \caption{We test our HopRAG against a series of baselines on multiple datasets using GPT-4o and GPT-3.5-turbo as the inference model with top 20 passages. We report the QA performance metrics EM and F1 score with GPT-3.5-turbo here and GPT-4o in Table \ref{tab:QA-performance-4o}, where the best score is in \textbf{bold} and the second best is \underline{underlined}.}
    \label{tab:QA-performance-3.5}
\end{table*}

\begin{table*}[htb]
    \centering
    \resizebox{0.7\textwidth}{!}{
    \begin{tabular}{ccccccccc}
        \toprule
        & \multicolumn{2}{c}{MuSiQue} & \multicolumn{2}{c}{2Wiki} & \multicolumn{2}{c}{HotpotQA} & \multicolumn{2}{c}{Average}\\
        
        Method& EM & F1 & EM & F1 & EM & F1  & EM &F1  \\
        \midrule
        BM25& 13.80 & 21.50 
& 40.30 & 44.83 
& 41.20 & 53.23 
 & 31.77 &39.85 \\
        BGE& 20.80 & 30.10 & 40.10 & 44.96 & 47.60 & 60.36  & 36.17 &45.14 \\
 \thead{Query \\ Decomposition}& 29.00& 38.50& 55.70& 60.57& 52.80& 68.67& 47.46&55.91\\
 Reranking& 32.00 & 40.29 & 53.70 & 58.44 & 55.40 & 70.03 & 48.61 &56.25 \\
 GraphRAG & 12.10 & 20.22 & 22.50 & 27.49 & 31.70 &42.74 
 & 22.10 &30.15 
\\
 RAPTOR & 36.40 & 49.09 & 53.80 & 61.45 & 58.00 &73.08  & 49.40 &61.21 
\\
 SiReRAG& \underline{40.50}& \underline{53.08}& \underline{59.60}& \underline{67.94}& \underline{61.70}&\textbf{76.48} & \underline{53.93}&\underline{65.83}\\ 
        \midrule
        HopRAG&  \textbf{42.20}&  \textbf{54.90}&  \textbf{61.10}&  \textbf{68.26}&  \textbf{62.00}&  \underline{76.06} & \textbf{55.10}&\textbf{66.40}\\
        \bottomrule
    \end{tabular}
    }
    \caption{We report the QA performance metrics EM and F1 score with GPT-4o and top 20 passages here,  where the best score is in \textbf{bold} and the second best is \underline{underlined}.}
    \label{tab:QA-performance-4o}
\end{table*}

\label{sec:experiment}
\subsection{Experimental Setups}
\paragraph{Datasets}
We collect several multi-hop QA datasets to evaluate the performance of HopRAG. We use HotpotQA dataset \citep{yang2018hotpotqadatasetdiverseexplainable}, 2WikiMultiHopQA dataset \citep{ho2020constructingmultihopqadataset} and MuSiQue dataset  \citep{trivedi2022musiquemultihopquestionssinglehop}. Following the same procedure as \citep{zhang2024sireragindexingsimilarrelated}, we obtain 1000 questions from each validation set of these three datasets. See Appendix \ref{dataset} for details.
\paragraph{Baselines}
We compare HopRAG with a variety of baselines: (1) unstructured RAG - sparse retriever BM25 \citep{robertson_probabilistic_2009} 
 (2) unstructured RAG - dense retriever BGE \citep{BGE,karpukhin2020densepassageretrievalopendomain} (3) unstructured RAG - dense retriever BGE with query decomposition \citep{min-etal-2019-multi} (4) unstructured RAG - dense retriever BGE with reranking \citep{nogueira2020passagererankingbert} (5) tree-structured RAG - RAPTOR \citep{sarthi_raptor_2024} (6) tree-structured RAG - SiReRAG \citep{zhang2024sireragindexingsimilarrelated} (7) graph-structured RAG - GraphRAG \citep{edge_local_2024} with the local search function (8) graph-structured RAG - HippoRAG \citep{gutiérrez2025hipporagneurobiologicallyinspiredlongterm}. For structured RAG baselines, we follow the same setting as previous work \citep{zhang2024sireragindexingsimilarrelated}. 

\paragraph{Metrics}
To measure the answer quality of different methods, we adopt exact match (EM) and F1 score which focus on the accuracy between a generated answer and the corresponding ground truth. We also use retrieval metrics to compare graph-based methods. Since tree-based methods like SiReRAG \citep{zhang2024sireragindexingsimilarrelated} and RAPTOR \citep{sarthi_raptor_2024} create new candidates (e.g., summary nodes) in the retrieval pool, it would be unfair to use retrieval metrics to compare them with others. We report both the answer and retrieval metrics in the ablations and discussion on HopRAG. See Appendix \ref{metric} for more metric details.
\paragraph{Settings}
We use BGE embedding model for semantic vectors at 768 dimensions. To avoid the loss of semantic information caused by chunking at a fixed size, we adopt the same chunking methods utilized in the original datasets respectively. 
GPT-4o-mini serves as both the model generating in-coming and out-coming questions when constructing the graph index, and the reasoning model for graph traversal. We use two reader models GPT-4o and GPT-3.5-turbo to generate the response given the context with 20 retrieval candidates and $n_{hop}=4$. See Appendix \ref{setup} for more setting details. 
\subsection{Main Results}
\begin{table*}[htb]
\centering
\setlength{\tabcolsep}{3pt}
\resizebox{0.9\textwidth}{!}{
      \begin{tabular}{ccccccccccccc}
    \toprule
 & \multicolumn{3}{c}{MuSiQue}& \multicolumn{3}{c}{2Wiki}& \multicolumn{3}{c}{HotpotQA} & \multicolumn{3}{c}{Average}\\
 & \multicolumn{2}{c}{Answer}&Retrieval& \multicolumn{2}{c}{Answer}&Retrieval& \multicolumn{2}{c}{Answer}&Retrieval & \multicolumn{2}{c}{Answer}&Retrieval \\
        
        $top_k$& EM& F1 & F1 & EM& F1 & F1 
& EM& F1 &F1 
 & EM& F1 &F1 
\\
        \midrule
        2& 32.50 & 46.31 & \textbf{37.83}&  47.80& 53.91 & \textbf{36.77}& 52.00& 67.78 &\textbf{50.23} 
 &  44.10& 56.00 &\textbf{41.61}\\
        4& 36.50 & 49.53 & \underline{35.02}& 54.50& 59.35 & \underline{33.22}& 55.60& 71.10 &\underline{46.45} 
 & 48.87& 59.99 &\underline{38.23}\\
 8& \underline{38.50}& 50.81 & 26.36 
& 56.10& 61.81 & 23.90 & 58.20& 75.05 &34.14  
 & 50.93 & 62.56 &28.13 \\
 12& 37.50 & \underline{51.47}& 
20.38 
& 57.70& 64.33 & 18.54 & \underline{59.50}& 75.54 &26.34 
 & 51.57& 63.78 &21.75 \\
 16& 37.50 & 51.44 & 16.47 
&  
\underline{60.00}& \underline{67.52}& 15.02 &  
\underline{59.50}& \underline{76.45}&21.75 
 & \underline{52.33}& \underline{65.14}&17.75 \\ 
        20&  \textbf{39.10}&  \textbf{53.00}&  13.89 & \textbf{61.60}& \textbf{68.93}& 12.51 & \textbf{61.30}& \textbf{78.34}&18.48   & \textbf{54.00}& \textbf{66.76}&14.96 \\
        \bottomrule
    \end{tabular}
}
    \caption{We test the robustness w.r.t hyperparameter $top_k$ on HopRAG using GPT-3.5-turbo on multiple datasets. We vary $top_k$ from 2 to 20 and report both the answer and retrieval metrics, where the best score is in \textbf{bold} and the second best is \underline{underlined}.}
    \label{tab:Ablation-3.5}
\end{table*}

\begin{table*}[htb]
\setlength{\tabcolsep}{2pt}
\centering
\resizebox{0.9\textwidth}{!}{
      \begin{tabular}{ccccccccc}
    \toprule
 &\multicolumn{2}{c}{MuSiQue}&\multicolumn{2}{c}{2Wiki}&\multicolumn{2}{c}{HotpotQA}&\multicolumn{2}{c}{Average}\\
        
        $n_{hop}$& Retrieval F1&LLM Cost& Retrieval F1&LLM Cost&Retrieval F1&LLM Cost&Retrieval F1&LLM Cost\\
        \midrule
        1&  8.78 &\textbf{20.00}&  8.68 &\textbf{19.86}& 6.78 &\textbf{19.91}& 8.08 &\textbf{19.92}\\
        2&  11.86 &\underline{30.32}&  11.42 &\underline{31.52}& 15.13 &\underline{29.39}& 12.80 &\underline{30.41}\\
 3&  \underline{12.67}&37.28 &  \underline{11.97}&37.15 
& \underline{16.76}&33.35& \underline{13.80}&35.93 
\\ 
        4&  \textbf{13.89}&40.32 & \textbf{12.51}&40.12 &\textbf{18.48}&35.14&\textbf{14.96}&38.53 \\
        \bottomrule
    \end{tabular}
    }
    \caption{We test the effect of hyperparameter $n_{hop}$ on HopRAG using GPT-3.5-turbo on multiple datasets with top 20 passages. We vary $n_{hop}$ from 1 to 4 and report both the answer and retrieval metrics in Table \ref{tab:nhopall}, and report the retrieval metrics here. For retrieval metrics, we calculate the retrieval F1 score and also the average number of calling LLM during traversal to measure the cost (the lower, the better). The best score is in \textbf{bold} and the second best is \underline{underlined}.}
    \label{tab:Ablation-3.5-hop}
\end{table*}

The main results are presented in Table \ref{tab:QA-performance-3.5} and \ref{tab:QA-performance-4o}. We observe that almost in all the settings HopRAG gives the best performance, with exceptions on HotpotQA when compared against SiReRAG and 2WikiMultiHopQA against HippoRAG. Overall, HopRAG achieves approximately 76.78\% higher than dense retriever (BGE), 48.62\% higher than query decomposition, 36.25\% higher than reranking (BGE), 9.94\% higher than RAPTOR,  3.08\% higher than HippoRAG,  1.11\% higher than SiReRAG. This illustrates the strengths of HopRAG in capturing both textual similarity and logical relations for handling multi-hop QA. 

Specifically, BM25, BGE and BGE with query decomposition yield unsatisfactory results since they rely solely on similarity, and BGE with reranking cannot capture logical relevance among candidates. Since GraphRAG considers relevance among entities instead of similarity for graph search, and RAPTOR focuses on the hierarchical logical relations among passages but cannot capture other kinds of relevance, both of them are more suitable for query-focused summarization but not the most competitive method for multi-hop QA tasks, as also reported in \citep{zhang2024sireragindexingsimilarrelated}.

\begin{table*}[htb]
\setlength{\tabcolsep}{3pt}
\centering
\resizebox{0.9\textwidth}{!}{
      \begin{tabular}{ccccccccccccc}
    \toprule
 & \multicolumn{3}{c}{MuSiQue}& \multicolumn{3}{c}{2Wiki}& \multicolumn{3}{c}{HotpotQA} & \multicolumn{3}{c}{Average}\\
 & \multicolumn{2}{c}{Answer}&Retrieval& \multicolumn{2}{c}{Answer}&Retrieval& \multicolumn{2}{c}{Answer}&Retrieval & \multicolumn{2}{c}{Answer}&Retrieval \\
        
        \thead{Method (Traversal Model)}& EM& F1 & F1 & EM& F1 & F1 
& EM& F1 &F1 
 & EM& F1 &F1 
\\
        \midrule
        BM25& 5.80& 11.00 
& 5.79& 27.00 & 31.55 & 9.25& 33.40& 44.30&8.75&  22.07 & 28.95 &7.93 \\
        BGE& 
11.80& 18.60& 8.76& 27.90 & 30.80 & 7.60 & 38.40& 50.56&11.10& 26.03 & 33.32 &9.16 \\
 \thead{HopRAG (non-LLM)}& 
19.00& 27.68& 8.27& 42.20& 46.72& 8.09 & 46.90& 61.17&11.73& 36.04& 45.19&9.36\\
 \thead{HopRAG (Qwen2.5-1.5B-Instruct)}& \underline{38.00}& \underline{46.73}& \underline{11.91}& \underline{58.40}& \underline{64.78}& \underline{11.82}& \underline{58.20}& \underline{74.74}& \underline{18.22}& \underline{51.53}& \underline{62.08}&\underline{13.98}\\ 
        \thead{HopRAG (GPT-4o-mini)}&    \textbf{39.10}&  \textbf{53.00}&  \textbf{13.89}& \textbf{61.60}& \textbf{68.93}& \textbf{12.51}& \textbf{61.30}& \textbf{78.34}&\textbf{18.48}& \textbf{54.00}& \textbf{66.76}&\textbf{14.96 }\\
        \bottomrule
    \end{tabular}
}
    \caption{We conduct an ablation study on the reasoning model during traversal with GPT-3.5-turbo as the inference model and top 20 passages. We compare 5 scenarios including sparse retriever (BM25), dense retriever (BGE), HopRAG (non-LLM), HopRAG (Qwen2.5-1.5B-Instruct) and HopRAG (GPT-4o-mini) and report both the answer and the retrieval metrics, where the best score is in \textbf{bold} and the second best is \underline{underlined}.}
    \label{tab:nonllm}
\end{table*}

In terms of HippoRAG, it prioritizes relevance signals such as vertices with the most edges and does not explicitly model similarity while our design HopRAG directly integrates similarity with logical relations when constructing edges. Although HopRAG only outperforms SiReRAG by a small margin in the scenario with top 20 candidate passages, our general graph structure does not introduce additional summary and proposition aggregate nodes and can facilitate efficient graph traversal for faster retrieval compared with SiReRAG. In the discussion, we will show that HopRAG can achieve competitive results with a smaller context length. Besides quantitative scores, we also demonstrate a case study in Appendix \ref{case study} comparing HopRAG and GraphRAG.
\subsection{Ablations and Discussion}
\label{ablation}
To confirm the robustness of HopRAG and provide more insights, we vary $top_k$, $n_{hop}$ and conduct ablation studies on traversal model. 
\paragraph{Effects of $top_k$} To show our efficiency in faster hop from indirectly relevant passages to truly relevant ones, we test the robustness by evaluating both the QA and retrieval performance on GPT-3.5-turbo with smaller $top_k$, as is shown in Table \ref{tab:Ablation-3.5}. From the results, we find that even with top 12 candidates, the QA performance of HopRAG is still comparable to that of HippoRAG or RAPTOR with 20 candidates, which highlights the effectiveness of our graph traversal design in efficiently retrieving more information within a limited context length. 
Meanwhile, we also observe that as $top_k$ increases, the retrieval F1 score gradually decreases due to the inclusion of excessive redundant information. Conversely, the answer quality generally improves, attributed to GPT-3.5-turbo's strong capability in processing and reasoning over extended contexts, with only one exception in the MuSiQue dataset.

\paragraph{Effects of $n_{hop}$} To assess the effects of the hyperparameter $n_{hop}$ on reasoning-augmented graph traversal, we vary $n_{hop}$ from 1 to 4 and evaluate the corresponding retrieval performance and cost, which is measured by the total number of LLM calls during graph traversal. The results shown in Table \ref{tab:Ablation-3.5-hop} indicate that as $n_{hop}$ increases, retrieval performance tends to improve, as more vertices are visited during traversal for reasoning and pruning. However, the expense and latency from calling LLM also increase with $n_{hop}$, creating a trade-off between performance and cost. We notice that as $n_{hop}$ increases, the number of new vertices in $C_{queue}$ requiring LLM reasoning decays rapidly. Since different vertices may hop to the same important vertex, the actual queue length in each round of hop is less than $top_k$. Specifically, the average queue length is 2.60 in the fourth round and 1.23 in the fifth round, suggesting that for the three datasets, the local area in the graph structure can be largely explored within four rounds of hop, eliminating the need for an additional hop. We set $n_{hop}$ as 4 in Table \ref{tab:QA-performance-3.5} and \ref{tab:QA-performance-4o}. We also evaluate the answer performances as $n_{hop}$ varies and show the overall results in Appendix \ref{nhop_ablation}.

\paragraph{Ablation on Traversal Model} 

In order to generalize HopRAG to scenarios with less computational overhead during retrieval, we supplement results from (1) HopRAG with traversal model Qwen2.5-1.5B-Instruct (2) HopRAG with non-LLM graph traversal that replaces the reasoning phase in Algorithm \ref{ragt} with similarity matching. Table \ref{tab:nonllm} shows that even without using the reasoning ability of LLM in the graph traversal, HopRAG can achieve 45.84\% higher than BM25 and 25.43\% higher than dense retriever (BGE), which proves the effectiveness of HopRAG in capturing textual similarity and logical relations for logic-aware retrieval.  The introduction of reasoning ability from LLM (GPT-4o-mini) can achieve about 45.78\% higher average score than the non-LLM version, and Qwen2.5-1.5B-Instruct as traversal model produces comparable results with less cost and higher efficiency. We analyze the retrieval efficiency in Appendix \ref{retrieval efficiency}.

\section{Conclusion}
\label{sec:conclusion}
In this paper, we introduced HopRAG, a novel RAG system with a logic-aware retrieval mechanism. HopRAG connects related passages through pseudo-queries, which allows identifying truly relevant passages within multi-hop neighborhoods of indirectly relevant ones, significantly enhancing both the precision and recall of retrieval.

Extensive experiments on multi-hop QA benchmarks, i.e. MuSiQue, 2WikiMultiHopQA, and HotpotQA, demonstrate that HopRAG outperforms conventional RAG systems and state-of-the-art baselines. Specifically, HopRAG achieved over 36.25\% higher answer accuracy and 20.97\% improved retrieval F1 score compared to conventional information retrieval approaches. It highlights the effectiveness of integrating logical reasoning into the retrieval module. Moreover, ablation studies provide insights into the sensitivity of hyperparameters and models, revealing trade-offs between retrieval performance and computational costs. 

HopRAG paves the way toward reasoning-driven knowledge retrieval. \textbf{Future work} involves scaling HopRAG to broader domains beyond QA tasks; optimizing indexing and traversal strategies for more complex scenarios with lower computation costs.

\section*{Acknowledgments}
This work is supported by the National Key R\&D Program of China (2024YFA1014003), National Natural Science Foundation of China (92470121, 62402016), CAAI-Ant Group Research Fund, and High-performance Computing Platform of Peking University.

\section*{Limitations}
Despite the benefits of HopRAG, the current evaluation focuses on multi-hop or multi-document QA tasks. In order to mitigate the risk of performance fluctuations when applying HopRAG to other datasets, we should explore its generalization capabilities across a broader range of domains. Besides, more sophisticated query simulation and edge merging strategies may lead to further improvements. Finally, though we were inspired by the theories of six degrees of separation and small-world networks, the degree distribution of our passage graph vertices does not exhibit the power-law characteristic. On the one hand, these theories only serve as the motivation and intuitive analogy; on the other hand, exploring more appropriate degree distribution strategies is an interesting research topic. We leave these research problems for future work.

\section*{Ethics Impact}
In our work, we acknowledge two key ethical considerations. First, we utilized AI assistant to enhance the writing process of our paper and code. We ensure that the AI assistant was used as a tool to improve clarity and conciseness, while the final content and ideas were developed and reviewed by human authors. Second, we employed multiple open source datasets and one open source tool Neo4j Community Edition \citep{10.1145/2384716.2384777} under GPL v3 license in our experiments. We are transparent about their origin and limitations, and we respect data ownership and user privacy.

\bibliography{acl_latex}
\newpage
\appendix
\section{Appendix}
\label{sec:appendix}
\localtableofcontents
\subsection{Symbols}
The symbols and their corresponding meanings are listed in Table \ref{tab:symbol}.
\begin{table*}[ht]
\setlength{\tabcolsep}{3pt}
\centering
\resizebox{0.8\textwidth}{!}{
    \begin{tabular}{cc|cc}
        \toprule
        symbol& meaning & symbol& meaning \\
        \midrule
        $P$& passage corpus& $Q_i^+$& set of out-coming questions for $p_i$\\
        $p$& passage & 
$q_{i,j}^+$& the j-th out-coming question for $p_i$\\
 $q$& query
& $Q_i^-$& set of in-coming questions for $p_i$\\
  $C$&retrieval context& $q_{i,j}^-$& the j-th in-coming question for $p_i$\\
 $\mathcal{P}(\cdot|\cdot)$&LLM distribution& 
$K_i^+$& set of keywords for $Q_i^+$ 
\\
 $\mathcal{O}$&response for $q$& $k_{i,j}^+$& keywords for $q_{i,j}^+$\\
 $G$&graph&  

$K_i^-$& set of keywords for $Q_i^-$ 
\\
  $\mathcal{V}$&set of vertices& $k_{i,j}^-$& keywords for $q_{i,j}^-$\\
  $v$&vertex&  
$V_i^+$& set of embeddings for $Q_i^+$ 
\\
 $\mathcal{E}$& set of directed edges& $\mathbf{v}_{i,j}^+$& embedding for $q_{i,j}^+$\\
 $e_{i,j}$& directed edge from $v_i$ to $v_j$& $V_i^-$& set of embeddings for $Q_i^-$ 
\\
 
$k_q$& keywords for $q$& $\mathbf{v}_{i,j}^-$& embedding for $q_{i,j}^-$\\
 $\mathbf{v}_q$& embedding for $q$& $R^+_i$& set of out-coming triplets\\
 $C_{count}$& counter of vertices during traversal& $r_{i,j}^+$&out-coming
 triplet for $q_{i,j}^+$\\
 $C_{queue}$& queue of vertices during traversal& $R^-_i$&set of in-coming triplets\\
 $H$& helpfulness metric& $r_{i,j}^-$&in-coming triplet for $q_{i,j}^-$\\
 
$top_k$& context budget& $n_{hop}$& number of hop\\
\bottomrule
    \end{tabular}
}
    \caption{Table of symbols and meanings.}
    \label{tab:symbol}
\end{table*}
\subsection{Datasets}
\label{dataset}
Table \ref{tab:dataset_statistics} shows the basic statistics of our datasets with their corresponding passage pool. Compared with knowledge graph, our graph-structured index is less dense and more efficient to construct. Since we put each passage text in the vertex, we can use fewer vertices to cover all the passages, which lowers the space complexity of the database. The average number of directed edges for each vertex is only 5.87, which lowers the time complexity for graph traversal. 
\begin{table*}[ht]
    \centering
    \setlength{\tabcolsep}{2pt}
    \begin{tabular}{cccccccc}
        \toprule
        dataset& number& docs &supporting facts& vertices& edges&  
avg text 
 length
& avg edge number\\
        \midrule
        MuSiQue & 1000 &   19990 &2800 &  13086 &  81348 
&  489.52 &  6.22 
\\
        2Wiki &  1000 &   10000 
&2388 &  23360 &  167068 
&  116.07 &  7.15 
\\
        HotpotQA & 1000 &   9942 
&2458 &  40534 &  171946 
&  132.44 &  4.24 
\\
        Average & 1000 &   13311 &2549 &  25660 &  140121 &  246.01 &  5.87 \\
        \bottomrule
    \end{tabular}
    \caption{Dataset Statistics. We report the basic statistics of the graph structure on different datasets and demonstrate that our efficient graph structure is traversal-friendly.}
    \label{tab:dataset_statistics}
\end{table*}
\subsection{Metrics}
\label{metric}

In our experiment, we mainly report the answer exact match (EM) and F1 score to compare all the methods.

     \paragraph{Exact Match (EM)} 
    The Exact Match (EM) metric measures the percentage of predictions that match any one of the ground truth answers exactly. It is defined as:
    \[
    EM = \frac{|\{ p \mid p = g \} |}{| P |}
    \]
    where \( p \) denotes a predicted answer, \( g \) denotes the corresponding ground truth answer, and \( P \) is the set of all the predictions.

     \paragraph{F1 Score}
    The F1 score is the harmonic mean of precision and recall, which measures the average overlap between the prediction and ground truth answer. Precision \( P \) and recall \( R \) are defined as:
    \[
    \text{P} = \frac{| A \cap \hat{A} |}{| \hat{A} |}, \quad \text{R} = \frac{| A \cap \hat{A} |}{| A |}
    \]
    where \( | A \cap \hat{A} | \) refers to the number of matching tokens between the prediction \( \hat{A} \) and the ground truth \( A \), and \( |\hat{A}| \), \( |A| \) denote the number of tokens in the predicted and ground truth answers, respectively.

    The F1 score is then computed as:
    \[
    F1 = \frac{2 \cdot \text{P} \cdot \text{R}}{\text{P} + \text{R}}
    \]

In our ablation study, we also report the retrieval F1 score to test the sensitivity of HopRAG, which is calculated as follows.

The Precision (P) and Recall (R) for retrieval are computed as:
\[
P = \frac{| \text{Ret} \cap \text{Rel} |}{| \text{Ret} |}, \quad R = \frac{| \text{Ret} \cap \text{Rel} |}{| \text{Rel} |}
\]
where \( \text{Ret} \) represents the set of passages retrieved during retrieval, and \( \text{Rel} \) denotes the set of relevant passages that support the ground truth answer.

The Retrieval F1 score is then calculated as the harmonic mean of precision and recall:
\[
F1_{\text{retrieval}} = \frac{2 \cdot P \cdot R}{P + R}
\]
\subsection{Settings}  
\label{setup}
To avoid semantic loss by chunking the documents at a fixed size, we chunk each document in a way corresponding to the supporting facts of each dataset. Specifically, we chunk each document in HotpotQA and 2WikiMultiHopQA by sentence since the smallest unit of these two datasets' supporting facts is a sentence.
To get embedding representation for each chunk, we use bge-base model. To extract keywords, we use the part-of-speech tagging function of Python package PaddleNLP to extract and filter entities.
In our method, we use the Neo4j graph database to store vertices and build edges. When building edges we employ prompt engineering technique to instruct the LLM to generate an appropriate number of questions for each vertex to cover its information, with a minimum requirement of at least 2 in-coming questions and 4 out-coming questions. To prevent the graph structure from becoming overly complex and dense, we retain only $O(n\cdot\log(n))$ edges, where $n$ is the number of vertices. We use GPT-4o-mini for reasoning-augmented graph traversal, GPT-4o and GPT-3.5-turbo for inference with 2048 max tokens and 0.1 temperature in our main experiments. 

For sparse and dense retrievers, we use the Neo4j database to conduct retrieval on the vertices. With this setting, we align the retrieval engine for unstructured baselines with HopRAG to fairly demonstrate the effectiveness of our graph structure index. For query decomposition, we use GPT-4o-mini to break down the query into multiple sub-queries, each of which should be a single-hop query. With $m$ sub-queries we conduct dense retrieval with BGE for each sub-query to get $top_k /m$ candidates independently and combine them all for final context. For reranking baseline, we use bge-reranker-base to rank $2*top_k$ candidates from dense retriever BGE and keep the $top_k$ ones as the final context. The structured baseline methods rely on specific open-source projects according to their papers.
\subsection{Prompts}
\label{prompt}
\begin{figure*}[t]
\centering
  \includegraphics[width=0.75\linewidth]{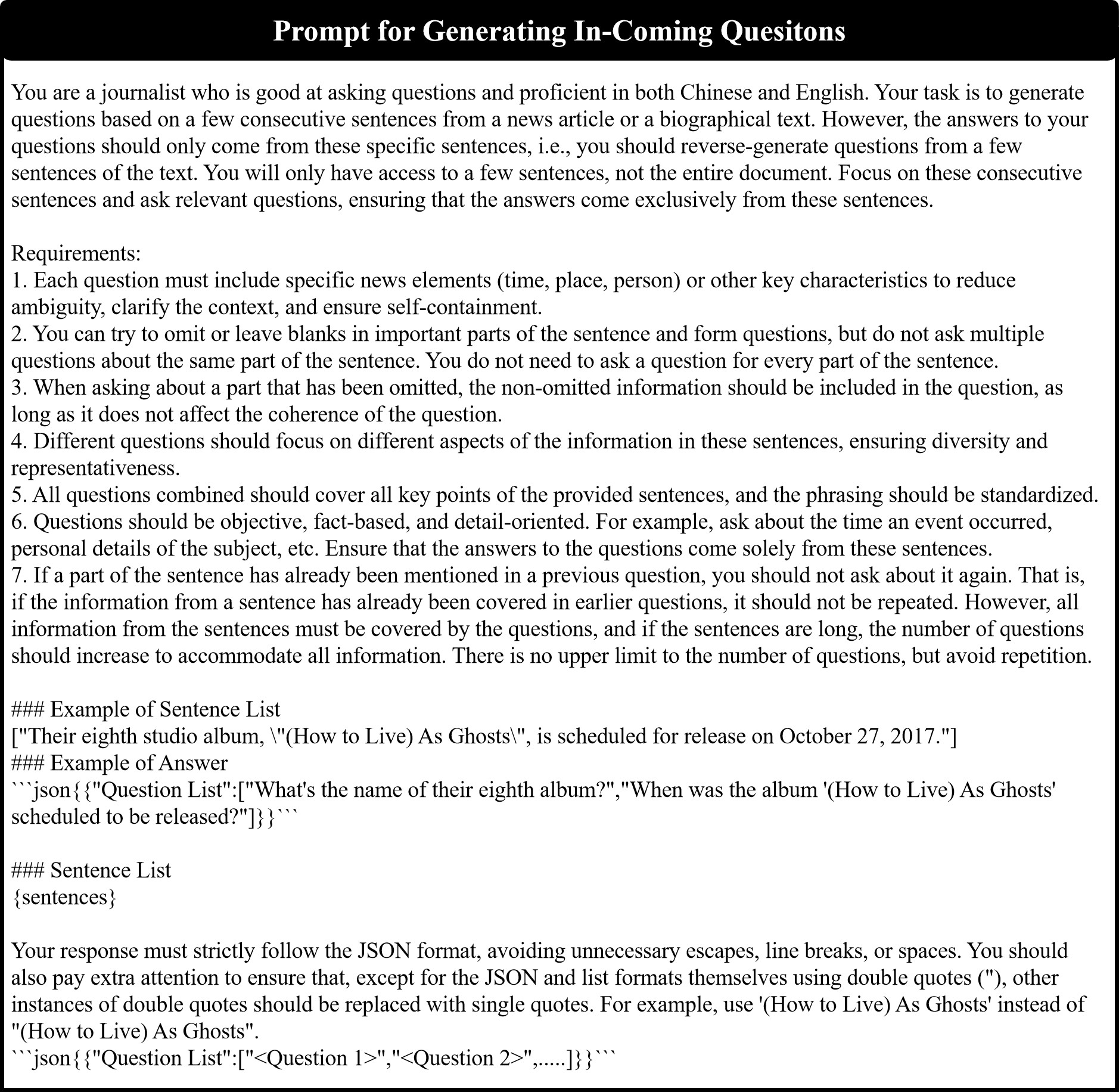}
  \caption{Prompt for generating in-coming questions.}
  \label{fig:incoming}
\end{figure*}

\begin{figure*}[t]
\centering
  \includegraphics[width=0.75\linewidth]{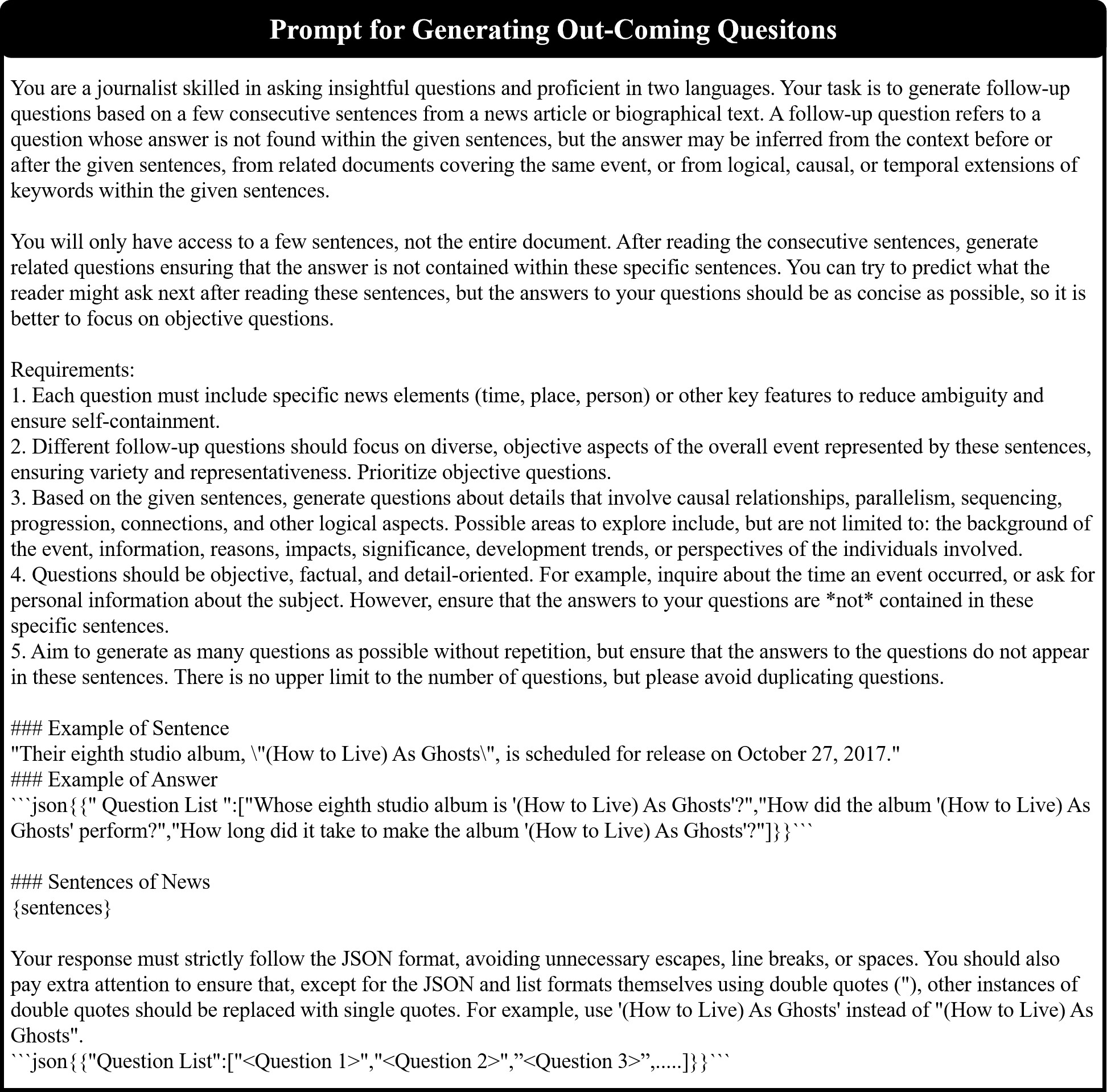}
  \caption{Prompt for generating out-coming questions.}
  \label{fig:outcoming}
\end{figure*}

\begin{figure*}[t]
\centering
  \includegraphics[width=1\linewidth]{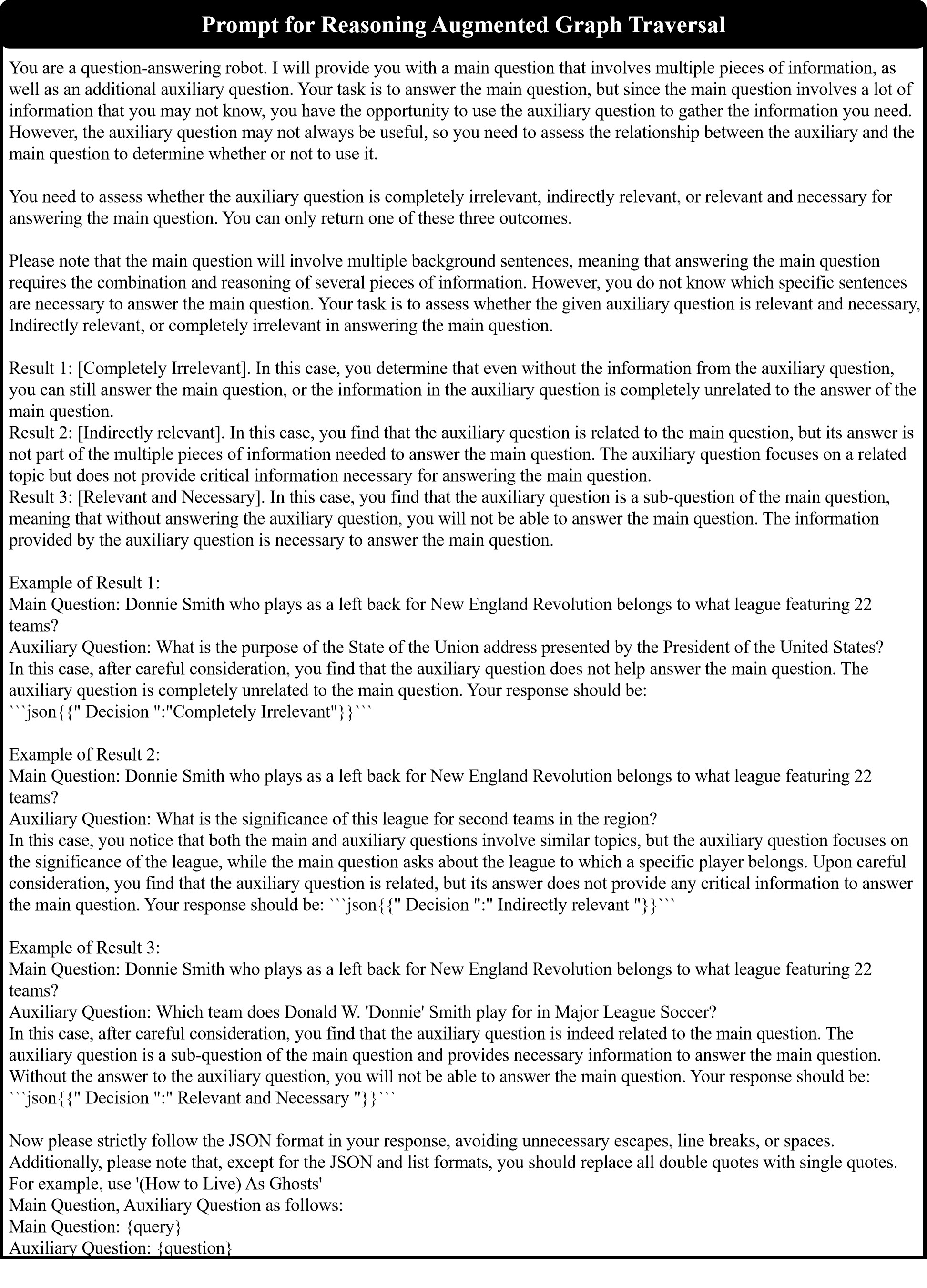}
  \caption{Prompt for reasoning-augmented graph traversal.}
  \label{fig:ragtprompt}
\end{figure*}

The prompt used for generating in-coming questions is shown in Figure \ref{fig:incoming}.  The prompt used for generating out-coming questions is shown in Figure \ref{fig:outcoming}. 
The prompt for reasoning-augmented graph traversal is shown in Figure \ref{fig:ragtprompt}. 

\begin{figure*}[ht]
\centering
\subfigure[Demonstration of one edge between two vertices.]{\label{fig:vertex_edge}\includegraphics[width=7cm,height=2cm]{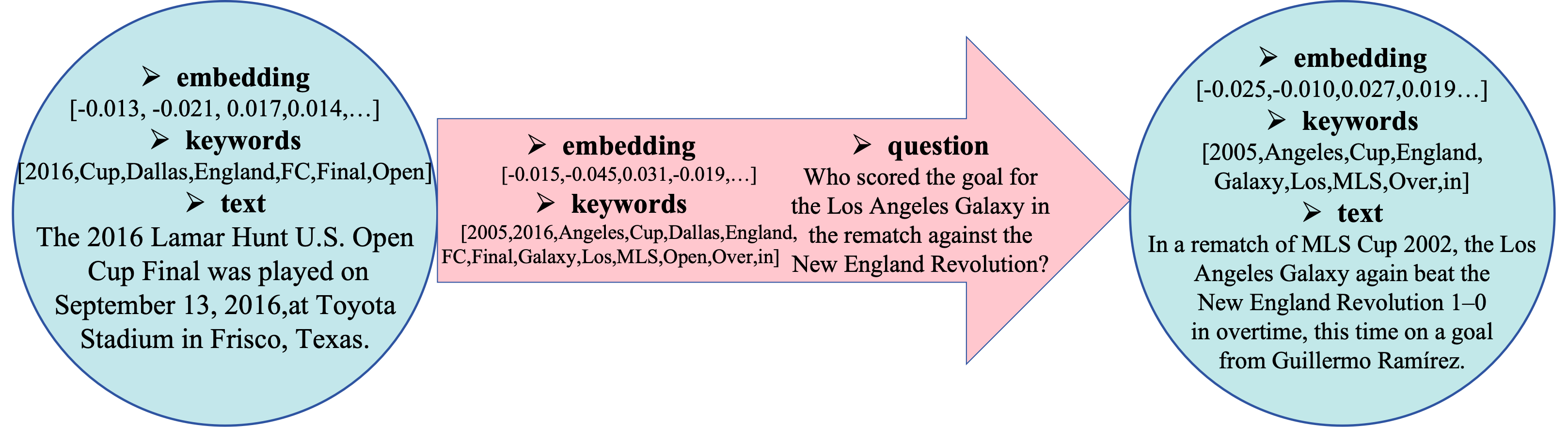}}
\subfigure[Demonstration of reasoning-augmented graph traversal.]
{\label{donnie demo}\includegraphics[width=4cm]{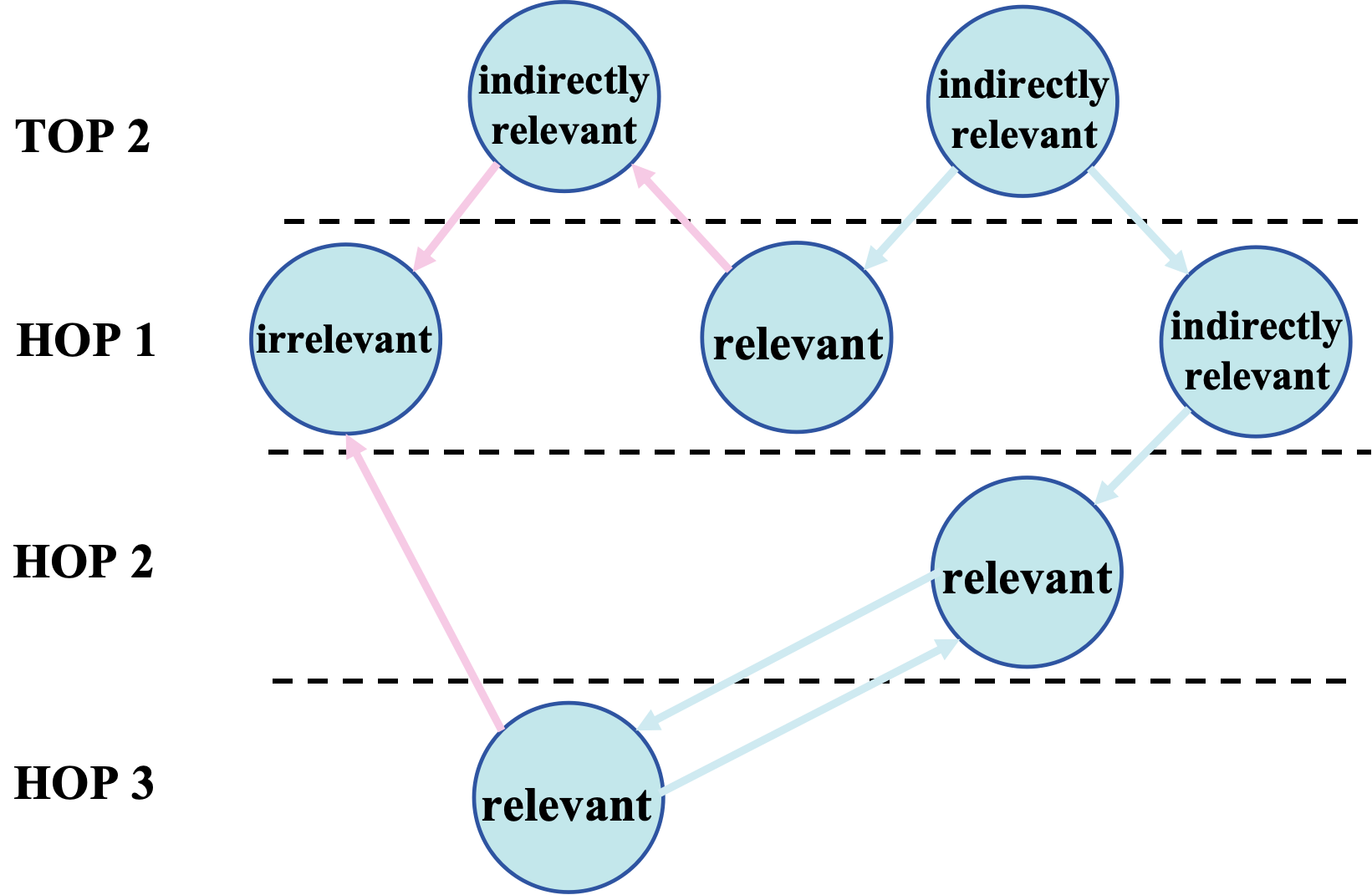}}
\caption{Demonstration of the traversal in the graph structure}\label{demo1}
\end{figure*}

\begin{figure*}[ht]
\centering
\subfigure[Demonstration of HopRAG's graph.]{\label{graph demo}\includegraphics[width=6cm,height=5cm]{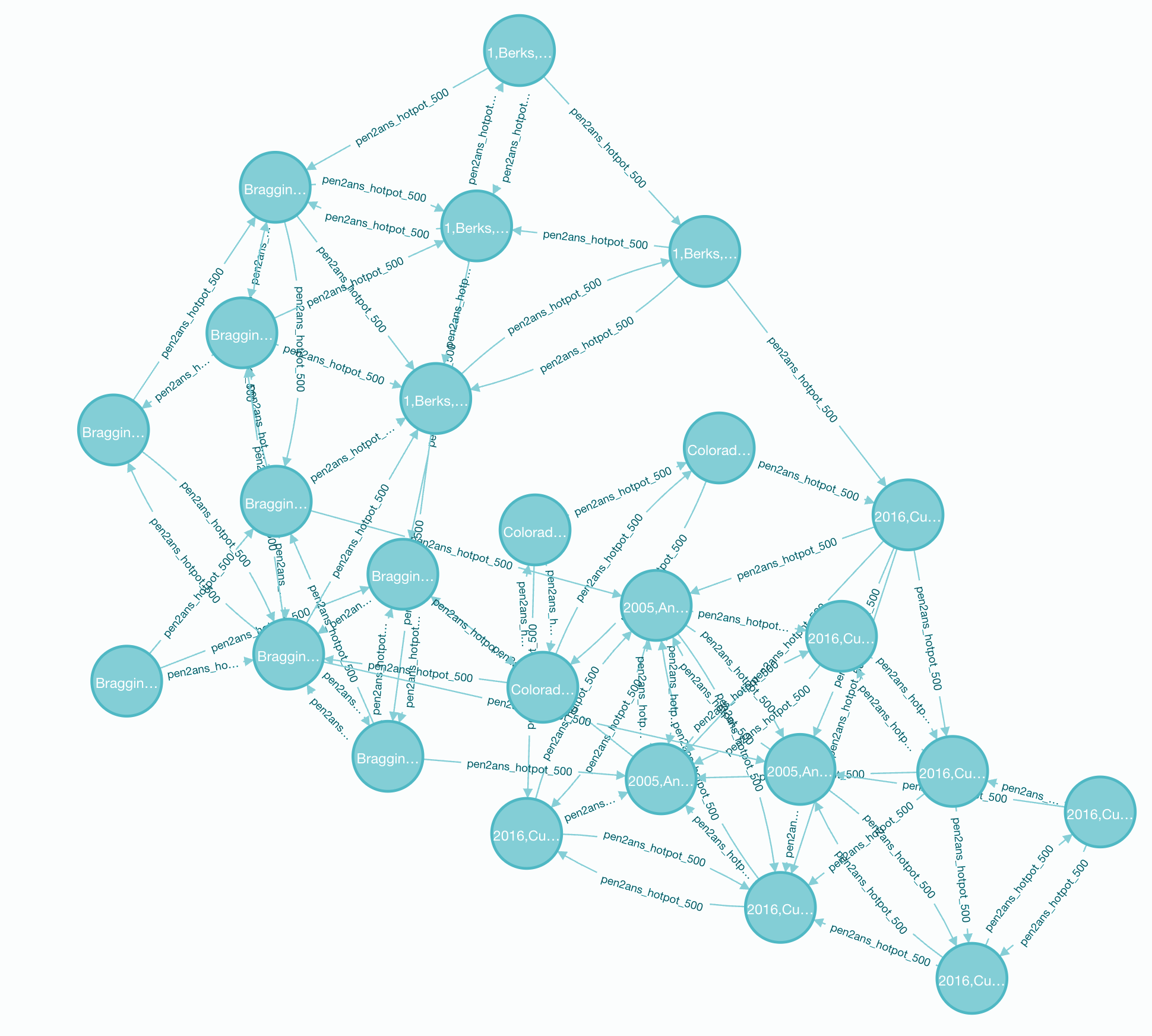}}
\subfigure[Demonstration of GraphRAG's graph.]
{\label{graphrag demo}\includegraphics[width=6cm]{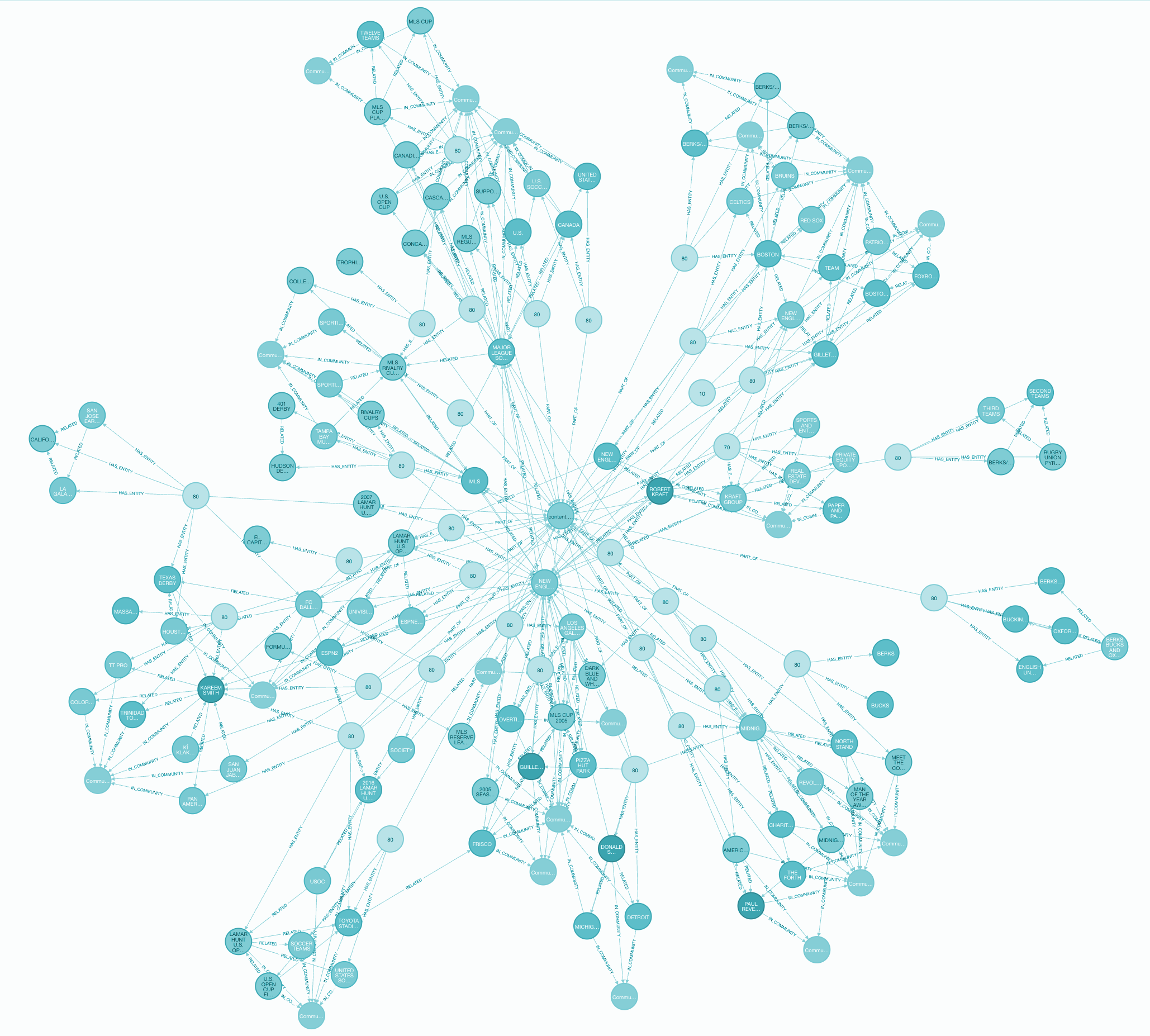}}
\caption{Visualizations of HopRAG and GraphRAG}\label{hoprag_graphrag_demo}
\end{figure*}

\subsection{Case Study}
\label{case study}
We demonstrate the graph structure in Figure \ref{graph demo}, one example edge with two vertices in Figure \ref{fig:vertex_edge}. Using the query "Donnie Smith who plays as a left back for New England Revolution belongs to what league featuring 22 teams?" as an example we conduct a qualitative analysis. For this multi-hop question (correct answer: Major League Soccer), the HotpotQA corpus contains three relevant sentences: (1) "Donald W. Donnie Smith (born December 7, 1990 in Detroit, Michigan) is an American soccer player who plays as a left back for New England Revolution in Major League Soccer."; (2) "Major League Soccer (MLS) is a men's professional soccer league, sanctioned by U.S. Soccer, that represents the sport's highest level in both the United States and Canada." and (3) "The league comprises 22 teams in the U.S. and 3 in Canada." 

With dense retriever (BGE), we can easily retrieve the first sentence but the last two facts can't be retrieved in the context even with a $top_k$ of 30. However, in our graph-structured index, these three vertices are logically connected, as is shown in Figure \ref{donnie demo}. During the traversal, LLM starts from the semantically similar but indirectly relevant vertices and can reach all the supporting facts within only a maximum of 3 hops.

Using this query and its passages for demonstration, we provide the visualizations of HopRAG’s graph index against GraphRAG to help readers grasp the differences and innovations. As shown in Figure  \ref{hoprag_graphrag_demo} , the main differences between HopRAG and GraphRAG are listed as follows.
\begin{itemize}
    \item \textbf{Vertices}. GraphRAG uses LLM to summarize the information from text chunks and then creates additional vertices (e.g., event, organization and person) in the graph, while HopRAG directly stores the original chunks in the vertices and thus avoids LLM hallucination during summarization, information loss during entity extraction and overly dense graph structure from redundant vertices. 
    \item \textbf{Edges}. GraphRAG connects vertices with pre-defined relationships like "part of" or "related", while HopRAG flexibly stores in-coming questions on the edges along with their keywords and embeddings, which can not only guide reasoning-augmented graph traversal but also facilitate edge retrieval.
    \item \textbf{Index}. GraphRAG creates and stores embeddings for the summarizations from LLM, while HopRAG creates sparse and dense indexes for both the vertices and the edges, which leads to more precise and efficient information retrieval.
\end{itemize}
In summary, the graph structure of HopRAG not only excavates logical relationships without creating additional vertices, but also paves the way for reasoning-driven knowledge retrieval. 

\begin{table*}[htb]
\setlength{\tabcolsep}{3pt}
\small
\centering
      \begin{tabular}{cccclccclccclcccl}
    \toprule
 & \multicolumn{4}{c}{MuSiQue}& \multicolumn{4}{c}{2Wiki}& \multicolumn{4}{c}{HotpotQA}& \multicolumn{4}{c}{Average}\\
 & \multicolumn{2}{c}{Answer}&\multicolumn{2}{c}{Retrieval}& \multicolumn{2}{c}{Answer}&\multicolumn{2}{c}{Retrieval}& \multicolumn{2}{c}{Answer}&\multicolumn{2}{c}{Retrieval}& \multicolumn{2}{c}{Answer}&\multicolumn{2}{c}{Retrieval}\\
        
        $n{hop}$& EM& F1 & F1  &Cost& EM& F1 & F1 
 &Cost& EM& F1 &F1 
  &Cost& EM& F1 &F1 
 &Cost\\
        \midrule
        1& 21.50 & 30.77 & 8.78 &\textbf{20.00}& 48.60 & 52.44 & 8.68 &\textbf{19.86}& 47.90& 62.92 &6.78 &\textbf{19.91}&  39.33& 48.71 &8.08 &\textbf{19.92}\\
        2& 
32.00 & 43.75 & 11.86 &\underline{30.32}& \underline{54.90}& \underline{60.37}& 11.42 &\underline{31.52}& 55.90 & 71.26 &15.13 &\underline{29.39}& \underline{47.60}& 58.46 &12.80 &\underline{30.41}\\
 3& 
\underline{32.50}& \underline{44.50}& \underline{12.67}&37.28 & 52.90& 59.16& \underline{11.97}&37.15 & \underline{57.40}& \underline{73.86}&\underline{16.76}&33.35 & \underline{47.60}& \underline{59.17}&\underline{13.80}&35.93 \\ 
        4&    \textbf{39.10}&  \textbf{53.00}&  \textbf{13.89}&40.32& \textbf{61.60}& \textbf{68.93}& \textbf{12.51}&40.12& \textbf{61.30}& \textbf{78.34}&\textbf{18.48}&35.14& \textbf{54.00}& \textbf{66.76}&\textbf{14.96}&38.53 \\
        \bottomrule
    \end{tabular}
    \caption{    
   We test the effect of hyperparameter $n_{hop}$ on HopRAG using GPT-3.5-turbo
with top 20 passages. We vary $n_{hop}$ from 1 to 4 and report both the answer and retrieval metrics. For answer metrics, we report the answer EM and F1 score; For retrieval metrics, we report the F1 score and average number of calling
LLM during traversal to measure the cost (the lower, the better). The best score is in \textbf{bold} and the second best is
\underline{underlined}. 
    }
    \label{tab:nhopall}
\end{table*}

\subsection{Retrieval Efficiency}
\label{retrieval efficiency}
We supplement more comprehensive analysis of retrieval efficiency, along with optimization strategies for further speedup. The main latency in HopRAG's retrieval comes from the LLM inference time during retrieval-augmented graph traversal. Since HopRAG with locally deployed Qwen2.5-1.5B-Instruct as the traversal model also showcases competitive performances, we focus on the retrieval efficiency in this scenario. Following the hyperparameters $n_{hop}=4$ and $top_k=20$ from our main experiments, each question requires calling LLM 38.53 times, where each LLM call involves selecting one edge from an average of 5.87 edges, with an input of around 500 tokens and an output of 20 tokens. According to \citep{Qwen25SpeedBenchmark}, the output token speed for Qwen2.5-1.5B-Instruct is about 40.86 token/s using BF16 and Transformer. Therefore, the additional latency for each question from LLM inference will be $38.53*20/40.86=18.86$ seconds. However, there are many optimization strategies to improve the retrieval efficiency. Using vLLM and GPTQ-Int4 techniques, the additional latency for each question can be reduced to $38.53*20/174.04=4.43$ seconds. Moreover, parallelism techniques like multithreading can further reduce the total execution time for all the queries.

\subsection{Discussion Results on \texorpdfstring{$n_{hop}$}{nhop}}
\label{nhop_ablation}

In the discussion we notice that as the hyper-parameter $n_{hop}$ varies from 1 to 4, the answer and retrieval performance both increase, along with the retrieval cost of calling LLM during traversal. Since the average queue length in the fifth hop is only as small as 1.23, we believe 4 is the ideal $n_{hop}$. The overall results are shown in Table \ref{tab:nhopall}.

\end{document}